\documentclass{kapproc} % Computer Modern font calls

\usepackage{xprocps}

\usepackage[dvips]{graphicx}

\upperandlowercase

\let\footnotetext\savefootnotetext

\let\footnoterule\savefootnoterule

\normallatexbib %will produce bibliography entries as shown in the
\begin{document}

\articletitle {Towards Sustainable Future by Transition to the Next
Level Civilisation\footnote{Report presented at the Symposium ``The
Future of Life and the Future of Our Civilisation'' (Frankfurt am
Main, 2--6 May 2005),
http://archive.future25.org/Symposium05/kirilyuk.pdf}}

\author{Andrei P. Kirilyuk\footnote{Address for
correspondence: Post Box 115, 01030 Kiev-30, Ukraine.}}

%% affil, email, and abstract are optional
\affil{Solid State Theory Department, Institute of Metal Physics \\
36 Vernadsky Avenue, 03142 Kiev-142, Ukraine}
\email{kiril@metfiz.freenet.kiev.ua}

% left running head:
\booktitlerunninghead {A.P.~Kirilyuk: Sustainable Future by
Transition to the Next Level Civilisation}

% right running head:
% \chaptitlerunninghead {Sustainable Future by Transition to the Next
% Level Civilisation}

\anxx{Kirilyuk\, Andrei P.}

\begin{abstract}
Universal and rigorously derived concept of dynamic complexity shows
that any system of interacting components, including society and
civilisation, is a process of highly uneven development of its
unreduced complexity. Modern civilisation state corresponds to the
end of unfolding of a big complexity level. Such exhausted, totally
``replete'' structure cannot be sustainable in principle and shows
instead increased instability, realising its replacement by a new
kind of structure with either low or much higher complexity
(degrading or progressive development branch respectively).
Unrestricted sustainability can emerge only after transition to the
next, superior level of civilisation complexity, which implies
qualitative and unified changes in all aspects of life, including
knowledge, production, social organisation, and infrastructure.
These changes are specified by a rigorous analysis of underlying
interaction processes. We propose mathematically rigorous
description of unreduced civilisation complexity development,
including universal criterion of progress. One obtains thus a
working basis for the causally complete, objectively exact and
reliable development science and futurology.
\end{abstract}

\begin{keywords}
Dynamic redundance, revolution of complexity, criterion of progress,
noosphere
\end{keywords}

\section{Future quest in a high-tech epoch of change}\label{sec:intro}
\markright {Sect. \ref{sec:intro}: Future quest in a high-tech epoch
of change}
Although permanent change is inherent in a planet, life, and
civilisation existence, it has a highly uneven character of
``punctuated equilibrium'', where larger periods of relatively
smooth and slow evolution are interrupted by short periods of huge
and abrupt, ``revolutionary'' change. Rapidly growing body of
evidence shows that today the planetary life and civilisation on
Earth are approaching very closely the next ``bifurcation point'' of
development, or ``generalised phase transition'' \cite{KirUSciCom},
which is often referred to as ``singularity'' (though in terms of
particular technological aspects) and marks a global change of
unprecedented scale (see e.g.
\cite{VernonVinge,ChangEarth,Rees,Soros,FutureShock}).

It is not surprising that the eternal humanity quest for its future
gains today quickly growing importance and public interest
\cite{WebFuture} that can be surpassed only by the global change
dynamics itself. A large part of this interest is driven by the
traditional ``fear of the (unknown) future'', essentially amplified
now because of the clearly felt huge scale of emerging change and
related uncertainty \cite{ChangEarth,Rees}. An important aspect of
the present epoch of change and its ``future shock''
\cite{FutureShock} is due to the extraordinary growth of ``high'',
but \emph{empirically} based technologies that can now, \emph{for
the first time in history}, modify the natural system complexity at
its \emph{full} depth, in quantum world (high-energy physics),
biology (genetics), environment (industrial over-production) and
human dimensions (psychology, media, information technologies),
while remaining \emph{effectively blind} at the level of genuine
\emph{understanding} of those real system dynamics
\cite{KirBlindTech3,KirBlindTech1,KirBlindTech2,KirBlindTech4}. Even
the most serious attempts of future studies \cite{WebFuture} fail to
provide an \emph{objectively} reliable, consistent and
\emph{unified} understanding of the emerging change meaning and
dynamics, replacing it with empirical \emph{interpolation} of
\emph{separate}, though important aspects of the current
development, such as economic and technological tendencies,
ecological system evolution, human behaviour, etc.

In this report we present the results of the \emph{causally
complete}, rigorous analysis of \emph{unreduced} planet and
civilisation dynamics based on the recently developed
\emph{universal concept of dynamic complexity}
\cite{KirUSciCom,KirBlindTech2,KirBlindTech4,KirSelfOrg,KirUSymCom}
and providing the \emph{unified}, many-sided picture, origin,
dynamics, and purpose of the beginning \emph{revolutionary} change
\cite{KirUSciCom,KirSustTrans}. We start with an outline of the
universal concept of complexity (Sec.~\ref{sec:USciCom}) emerging
from the \emph{unreduced} solution to any \emph{real} interaction
problem (Sec.~\ref{subsec:UnredInteract}) and leading to the unified
concept of system \emph{development} as manifestation of the
universal \emph{symmetry (conservation)} of complexity
(Sec.~\ref{subsec:Evolution}). In particular, the
\emph{sustainability transition} emerges today as inevitable and
rather rapid ``jump'' to the next, \emph{superior level of
civilisation complexity} (Sec.~\ref{sec:Transition}) prepared by all
its previous development and having only one alternative of
irreversible destruction (Sec.~\ref{subsec:TodayBifurc}). We then
analyse various entangled aspects of life at the new complexity
level and corresponding transition dynamics, including the
qualitatively new kind of knowledge (Sec.~\ref{subsec:NewScience}),
production (Sec.~\ref{subsec:NewProd}), social organisation
(Sec.~\ref{subsec:NewSocial}), and infrastructure
(Sec.~\ref{subsec:NewSettle}). Finally, we pay homage to Carl Sagan
and Joseph Shklovsky by showing that discovery of other forms of
life and intelligence is related to the new future for our own
civilisation by the same universal concept of complexity
(Sec.~\ref{sec:Conclusion}). We summarise the obtained results by
concluding that the \emph{causally complete} kind of knowledge of
the universal science of complexity provides the \emph{unique basis}
for the truly scientific, \emph{objectively reliable and
intrinsically unified futurology} urgently needed especially at the
modern critical point of development.

\section{Universal science of complexity}\label{sec:USciCom}
\markright {Sect. \ref{sec:USciCom}: Universal science of
complexity}
\subsection{Unreduced interaction dynamics}\label{subsec:UnredInteract}
Any system dynamics and evolution are determined by the underlying
interaction processes. The way of interaction analysis in usual
science (including the scholar ``science of complexity'') involves
rough simplification (reduction) of real interaction within a
version of perturbation theory (or ``model'') that assumes effective
\emph{weakness} of interaction influence upon system configuration,
which \emph{kills} any possibility of essential \emph{novelty
emergence} from the beginning (with the evident fatal consequences
for such approach ability to predict \emph{any} nontrivial future).
Subsequent play with analytical or computer models of thus heavily
reduced reality, empirically postulated (rather than derived) object
properties and arbitrarily adjusted parameters cannot replace the
\emph{intrinsic creativity} of unreduced interaction processes. It
is no wonder that the \emph{qualitative} knowledge extension to the
\emph{causally complete understanding} of real phenomena, provided
by the universal science of complexity \cite{KirUSciCom}, is simply
due to the proposed \emph{non-simplified, truly ``exact'' analysis
of unreduced, real interaction processes}. Its possibilities are
confirmed by the obtained \emph{consistent solutions} to various
stagnating, ``insoluble'' problems \cite{KirUSciCom}, from those of
fundamental physics (causal and unified extensions of quantum
mechanics, relativity, cosmology)
\cite{KirBlindTech1,KirQFM,KirCosmo} and unreduced many-body
interaction (true quantum chaos, quantum measurement, many-body
coherence) \cite{KirBlindTech2,KirQuChaos}, to reliable basis for
nanobiotechnology \cite{KirBlindTech2,KirNano}, genomics
\cite{KirBlindTech4} and medicine \cite{KirFractal}, theory of
genuine (natural or artificial) intelligence and consciousness
\cite{KirConscious}, the new kind of communication and information
systems \cite{KirCommNet}, and realistic sustainability concept
\cite{KirSustTrans}.

Any real interaction can be represented by \emph{existence
equation}, generalising various models and simply fixing the initial
system configuration in a ``Hamiltonian'' form (self-consistently
confirmed later)
\cite{KirUSciCom,KirBlindTech2,KirBlindTech4,KirSelfOrg,KirUSymCom,KirCosmo,KirConscious,KirCommNet}:
\begin{equation}\label{eq:1}
\left\{ {\sum\limits_{k = 0}^N {\left[ {h_k \left( {q_k } \right) +
\sum\limits_{l > k}^N {V_{kl} \left( {q_k ,q_l } \right)} } \right]}
} \right\}\Psi \left( Q \right) = E\Psi \left( Q \right)\ ,
\end{equation}
where $h_k \left( {q_k } \right)$ is the ``generalised Hamiltonian''
for the $k$-th component, $q_k$ is the degree(s) of freedom of the
$k$-th component, $V_{kl} \left( {q_k ,q_l } \right)$ is the
(arbitrary) interaction potential between the $k$-th and $l$-th
components, $\Psi \left( Q \right)$ is the system state-function, $Q
\equiv \left\{ {q_0,q_1,...,q_N } \right\}$, $E$ is the generalised
Hamiltonian eigenvalue, and summations include all ($N$) system
components. It is convenient to represent the same equation in
another form by separating certain degree(s) of freedom, e.g. $q_0
\equiv \xi $, that correspond to a naturally selected, usually
``system-wide'' entity, such as ``embedding'' configuration (system
of coordinates) or common ``transmitting agent'':
\begin{equation}\label{eq:2}
\left\{ {h_0 \left( \xi  \right) + \sum\limits_{k = 1}^N {\left[
{h_k \left( {q_k } \right) + V_{0k} \left( {\xi ,q_k } \right)} +
\sum\limits_{l > k}^N {V_{kl} \left( {q_k ,q_l } \right)} \right]} }
\right\}\Psi \left( {\xi ,Q} \right) = E\Psi \left( {\xi ,Q}
\right),
\end{equation}
where now $Q \equiv \left\{ {q_1 ,...,q_N } \right\}$ and $k,l \ge
1$.

We pass now to a ``natural'' problem expression in terms of
free-component solutions for the ``functional'' degrees of freedom
($k \ge 1$):
\begin{equation}\label{eq:3}
h_k \left( {q_k } \right)\varphi _{kn_k } \left( {q_k } \right) =
\varepsilon _{n_k } \varphi _{kn_k } \left( {q_k } \right)\ ,
\end{equation}
\begin{equation}\label{eq:4}
\Psi \left( {\xi ,Q} \right) = \sum\limits_n {\psi _n \left( \xi
\right)} \varphi _{1n_1 } \left( {q_1 } \right)\varphi _{2n_2 }
\left( {q_2 } \right)...\varphi _{Nn_N } \left( {q_N } \right)
\equiv \sum\limits_n {\psi _n \left( \xi  \right)} \Phi _n \left( Q
\right),
\end{equation}
where $\left\{ {\varepsilon _{n_k } } \right\}$ are the eigenvalues
and $\left\{ {\varphi _{kn_k } \left( {q_k } \right)} \right\}$
eigenfunctions of the $k$-th component Hamiltonian $h_k \left( {q_k
} \right)$, $n \equiv \left\{ {n_1,...,n_N } \right\}$ runs through
all eigenstate combinations, and $\Phi _n \left( Q \right) \equiv
\varphi _{1n_1 } \left( {q_1 } \right)\varphi _{2n_2 } \left( {q_2 }
\right)...\varphi _{Nn_N } \left( {q_N } \right)$ by definition. The
system of equations for $\left\{ {\psi _n \left( \xi  \right)}
\right\}$, equivalent to the starting existence equation
(\ref{eq:1})--(\ref{eq:2}) is obtained in a standard way
\cite{KirUSciCom,KirBlindTech2,KirBlindTech4,KirSelfOrg,KirUSymCom,KirConscious,KirCommNet}:
\begin{equation}\label{eq:5}
\begin{array}{lll}
\left[ {h_0 \left( \xi  \right) + V_{00} \left( \xi  \right)}
\right]\psi _0 \left( \xi  \right) + \sum\limits_n {V_{0n} \left(
\xi  \right)} \psi _n \left( \xi  \right) = \eta \psi _0 \left(
\xi  \right)\\
\left[ {h_0 \left( \xi  \right) + V_{nn} \left( \xi \right)}
\right]\psi _n \left( \xi  \right) + \sum\limits_{n' \ne n} {V_{nn'}
\left( \xi  \right)} \psi _{n'} \left( \xi  \right) = \eta _n \psi
_n \left( \xi  \right) - V_{n0} \left( \xi \right)\psi _0 \left( \xi
\right),
\end{array}
\end{equation}
where $n,n' \ne 0$ (also below), $\eta \equiv \eta _0 = E -
\varepsilon _0$, $\eta _n = E - \varepsilon _n$, $\varepsilon _n =
\sum\limits_k {\varepsilon _{n_k } }$,
\begin{equation}\label{eq:6}
V_{nn'} \left( \xi  \right) = \sum\limits_k {\left[ {V_{k0}^{nn'}
\left( \xi  \right) + \sum\limits_{l > k} {V_{kl}^{nn'} } }
\right]}\ ,
\end{equation}
\begin{equation}\label{eq:7}
V_{k0}^{nn'} \left( \xi  \right) = \int\limits_{\Omega _Q } {dQ\Phi
_n^ *  \left( Q \right)} V_{k0} \left( {q_k ,\xi } \right)\Phi _{n'}
\left( Q \right)\ ,
\end{equation}
\begin{equation}\label{eq:8}
V_{kl}^{nn'} \left( \xi \right) = \int\limits_{\Omega _Q } {dQ\Phi
_n^ *  \left( Q \right)} V_{kl} \left( {q_k ,q_l } \right)\Phi _{n'}
\left( Q \right)\ ,
\end{equation}
and we have separated the equation for $\psi _0 \left( \xi \right)$
describing the generalised ``ground state'' of the system elements,
i. e. the state with minimum complexity (defined below). The
obtained system of equations (\ref{eq:5}) expresses the same problem
as the starting Eq.~(\ref{eq:2}), but now in terms of intrinsic
variables. Therefore it can be obtained for various starting models,
including time-dependent and formally ``nonlinear'' ones.

The usual, perturbative approach starts from explicit simplification
of the ``nonintegrable'' system (\ref{eq:5}) down to a
``mean-field'' approximation:
\begin{equation}\label{eq:9}
\left[ {h_0 \left( \xi  \right) + V_{nn} \left( \xi  \right) +
\tilde V_n \left( \xi  \right)} \right]\psi _n \left( \xi  \right) =
\eta _n \psi _n \left( \xi  \right)\ ,
\end{equation}
where $ \left| {V_0 \left( \xi  \right)} \right|  < \left| {\tilde
V_n \left( \xi  \right)} \right| < \left| {\sum\limits_{n'} {V_{nn'}
} \left( \xi  \right)} \right| $. General problem solution is then
obtained as a linear or equivalent \emph{superposition} of
eigen-solutions of Eq.~(\ref{eq:9}) similar to Eq.~(\ref{eq:4}). If
we want to avoid problem reduction, we can try to ``solve'' the
unsolvable system (\ref{eq:5}) by expressing $\psi _n \left( \xi
\right)$ through $\psi _0 \left( \xi \right)$ from the equations for
$\psi _n \left( \xi  \right)$ using the standard Green function
technique and then substituting the result into the equation for
$\psi _0 \left( \xi  \right)$ \cite{KirChannel,Dederichs}. We are
left then with only one, formally ``integrable'' equation for $\psi
_0 \left( \xi \right)$:
\begin{equation}\label{eq:10}
h_0 \left( \xi  \right)\psi _0 \left( \xi  \right) + V_{{\rm{eff}}}
\left( {\xi ;\eta } \right)\psi _0 \left( \xi \right) = \eta \psi _0
\left( \xi  \right)\ ,
\end{equation}
where the operator of \emph{effective potential (EP)},
$V_{{\rm{eff}}} \left( {\xi ;\eta } \right)$, is obtained as
\begin{equation}\label{eq:11}
V_{{\rm{eff}}} \left( {\xi ;\eta } \right) = V_{00} \left( \xi
\right) + \hat V\left( {\xi ;\eta } \right),\ \  \hat V\left( {\xi
;\eta } \right)\psi _0 \left( \xi  \right) = \int\limits_{\Omega
_\xi  } {d\xi 'V\left( {\xi ,\xi ';\eta } \right)} \psi _0 \left(
{\xi '} \right),
\end{equation}
\begin{equation}\label{eq:12}
V\left( {\xi ,\xi ';\eta } \right) = \sum\limits_{n,i}
{\frac{{V_{0n} \left( \xi  \right)\psi _{ni}^0 \left( \xi
\right)V_{n0} \left( {\xi '} \right)\psi _{ni}^{0*} \left( {\xi '}
\right)}}{{\eta  - \eta _{ni}^0  - \varepsilon _{n0} }}}\ ,\ \ \
\varepsilon _{n0}  \equiv \varepsilon _n  - \varepsilon _0\ ,
\end{equation}
and $\left\{ {\psi _{ni}^0 \left( \xi  \right)} \right\}$, $\left\{
{\eta _{ni}^0 } \right\}$ are complete sets of eigenfunctions and
eigenvalues of a \emph{truncated} system of equations:
\begin{equation}\label{eq:13}
\left[ {h_0 \left( \xi  \right) + V_{nn} \left( \xi  \right)}
\right]\psi _n \left( \xi  \right) + \sum\limits_{n' \ne n} {V_{nn'}
\left( \xi  \right)} \psi _{n'} \left( \xi  \right) = \eta _n \psi
_n \left( \xi  \right)\ .
\end{equation}
The \emph{unreduced}, truly complete \emph{general solution} to a
problem emerges now as a \emph{dynamically probabilistic} sum of
\emph{redundant} system \emph{realisations}, each of them equivalent
to the whole usual ``general solution''
\cite{KirUSciCom,KirBlindTech2,KirBlindTech4,KirSelfOrg,KirUSymCom,KirQuChaos}:
\begin{equation}\label{eq:14}
\rho \left( {\xi ,Q} \right) = \sum\limits_{r = 1}^{N_\Re  } {^{^
\oplus}  \rho _r \left( {\xi ,Q} \right)}\ ,
\end{equation}
where $\rho \left( {\xi ,Q} \right)$ is the observed density, $\rho
\left( {\xi ,Q} \right) = \left| {\Psi \left( {\xi ,Q} \right)}
\right|^2 $ for ``wave-like'' complexity levels and $\rho \left(
{\xi ,Q} \right) = \Psi \left( {\xi ,Q} \right)$ for
``particle-like'' structures, index $r$ enumerates system
realisations, $N_\Re$ is realisation number (its maximum value is
equal to the number of components, $N_\Re = N$), and the sign
$\oplus$ designates the special, dynamically probabilistic meaning
of the sum (see below). The $r$-th realisation state-function, $\Psi
_r \left( {\xi ,Q} \right)$, entering the unreduced general
solution, Eq.~(\ref{eq:14}), is obtained as
\[
\Psi _r \left( {\xi ,Q} \right) = \sum\limits_i {c_i^r } \left[
{\Phi _0 \left( Q \right)\psi _{0i}^r \left( \xi  \right)} \right. +
\]
\begin{equation}\label{eq:15}
+ \sum\limits_{n, i'} {\left. {\frac{{\Phi _n \left( Q \right)\psi
_{ni'}^0 \left( \xi  \right)\int\limits_{\Omega _\xi } {d\xi '\psi
_{ni'}^{0*} \left( {\xi '} \right)V_{n0} \left( {\xi '} \right)\psi
_{0i}^r \left( {\xi '} \right)} }}{{\eta _i^r  - \eta _{ni'}^0  -
\varepsilon _{n0} }}} \right] }\ ,
\end{equation}
where $\{ \psi _{0i}^r \left( \xi  \right),\eta _i^r \}$ are
eigen-solutions of the unreduced EP equation (\ref{eq:10}), and the
$r$-th EP realisation takes the form:
\[
V_{{\rm{eff}}} \left( {\xi ;\eta _i^r } \right)\psi _{0i}^r \left(
\xi  \right) = V_{00} \left( \xi  \right)\psi _{0i}^r \left( \xi
\right) +
\]
\begin{equation}\label{eq:16}
+ \sum\limits_{n, i'} {\frac{{V_{0n} \left( \xi  \right)\psi
_{ni'}^0 \left( \xi  \right)\int\limits_{\Omega _\xi  } {d\xi '\psi
_{ni'}^{0*} \left( {\xi '} \right)V_{n0} \left( {\xi '} \right)\psi
_{0i}^r \left( {\xi '} \right)} }}{{\eta _i^r  - \eta _{ni'}^0  -
\varepsilon _{n0} }}}\ .
\end{equation}
Although the ``effective'' problem,
Eqs.~(\ref{eq:10})--(\ref{eq:16}), is formally equivalent to its
initial expression, Eqs.~(\ref{eq:1})--(\ref{eq:5}), it reveals
\emph{emerging interaction links}, in the form of EP dependence on
the solutions to be found. It leads to a \emph{new quality} of the
\emph{unreduced} solution (as compared to usual reduction of Eq.
(\ref{eq:9})): the former has \emph{many} equally real, locally
``complete'' and therefore mutually \emph{incompatible} solutions
called (system) \emph{realisations}
\cite{KirUSciCom,KirBlindTech2,KirBlindTech4,KirSelfOrg,KirUSymCom,KirCosmo,KirQuChaos,KirFractal,KirConscious,KirCommNet,KirChannel}.
This quality of the unreduced solution is designated as
\emph{dynamic multivaluedness (or redundance)}. Standard theory
tries to obtain problem solution in a ``closed'', ``exact'' form and
therefore resorts to perturbative reduction of the original EP (see
e.g. \cite{Dederichs}), thus inevitably \emph{killing} real system
multivaluedness, complexity and \emph{creativity}.

Dynamic multivaluedness gives \emph{dynamic, or causal, randomness}:
multiple, but incompatible system realisations are forced, by the
\emph{same} driving interaction, to \emph{permanently replace each
other} in a \emph{truly random} order (thus defined), which leads to
the \emph{unreduced} general solution in the form of (dynamically)
\emph{probabilistic sum} of Eq.~(\ref{eq:14}). It implies that any
quantity is \emph{intrinsically unstable} and its value will
\emph{unpredictably} change (together with the system state) to
another one, corresponding to the next, \emph{randomly} chosen
realisation. We obtain thus a \emph{consistently derived} and
\emph{universally valid} property of \emph{novelty emergence}, or
intrinsic \emph{creativity} of \emph{any} real system, absent in any
its usual, dynamically single-valued model. We obtain also purely
\emph{dynamic} definition of realisation emergence \emph{event} and
its \emph{probability}:
\begin{equation}\label{eq:17}
\alpha _r \left( {N_r } \right) = \frac{{N_r }}{{N_\Re  }}\ \ \
\left( {N_r  = 1,...,N_\Re  ;\ \sum\limits_r {N_r }  = N_\Re  }
\right),\ \ \ \sum\limits_r {\alpha _r }  = 1\ ,
\end{equation}
where $\alpha _r$ is the probability of $r$-th actually observed
realisation that contains $N_r$ elementary realisations ($N_r = 1$
for each of these).

The obtained picture of real system dynamics can be summarised by
the \emph{universal} definition of unreduced \emph{dynamic
complexity}, $C$, as any growing function of realisation number,
$N_\Re$, or rate of change, equal to zero for the (unrealistic) case
of only one realisation: $C = C\left( {N_\Re  } \right)$, ${{dC}
\mathord{\left/ {\vphantom {{dC} {dN_\Re > 0}}} \right.
\kern-\nulldelimiterspace} {dN_\Re
> 0}}$, $C\left( {\rm{1}} \right){\rm{  = 0}}$. Major examples
are provided by $C\left( {N_\Re  } \right) = C_0 \ln N_\Re$,
generalised energy/mass (temporal rate of realisation change), and
momentum (spatial rate of realisation emergence)
\cite{KirUSciCom,KirBlindTech2,KirBlindTech4,KirSelfOrg,KirUSymCom,KirQFM,KirCosmo,KirConscious,KirCommNet}.
Since dynamic redundance ($N_\Re > 1$) is at the origin of dynamic
randomness, our dynamic complexity includes universally defined
\emph{chaoticity}. Whereas \emph{all} real systems and processes are
dynamically complex and (internally) chaotic ($N_\Re > 1,\ C
> 0$), their ``models'' in usual science, including its
versions of ``complexity'' and ``chaoticity'' (cf.
\cite{Perplexity,EndScience}), are invariably produced by artificial
(and \emph{biggest} possible) reduction of multivalued dynamics to
the unrealistic case of \emph{single} realisation, \emph{zero}
complexity, \emph{absence} of \emph{genuine} chaos, \emph{any} real,
\emph{intrinsic} change and related \emph{time flow}. This
\emph{dynamically single-valued}, or \emph{unitary}, science
embracing the whole body of scholar knowledge is a
\emph{zero-dimensional (point-like) projection} of multivalued world
dynamics, which explains both relative (but never complete!)
``success'' of unitary science in its \emph{formal} description of
the \emph{lowest} complexity levels ($\simeq$ ``fundamental
physics'') and its explicit failure to understand higher-level
dynamics and unreduced complexity features (emergence, time, chaos,
etc.)
\cite{KirUSciCom,KirBlindTech2,KirBlindTech4,KirSelfOrg,KirUSymCom,KirCosmo,KirConscious}.

Unreduced dynamic complexity thus defined includes other major
features, such as essential (or dynamic) nonlinearity, dynamic
entanglement, and probabilistic dynamic fractality. \emph{Essential
nonlinearity} designates \emph{dynamically} emerging feedback links,
described by EP dependence on the eigenvalues to be found
(Eqs.~(\ref{eq:10})--(\ref{eq:12}),(\ref{eq:16})). It is only
incorrectly modelled by usual, mechanistic ``nonlinearity'' of the
unitary theory and appears in interaction problems with a formally
linear existence equation (\ref{eq:1})--(\ref{eq:2}), such as
quantum chaos \cite{KirQuChaos,KirChannel}. \emph{Dynamic
entanglement} is \emph{physically real} mixing of interacting
components reflected by the dynamically weighted products of
functions depending on different degrees of freedom in
Eq.~(\ref{eq:15}). Both essential nonlinearity and dynamic
entanglement are amplified due to multi-level realisation branching
giving \emph{probabilistic dynamical fractal}. It is obtained by
application of the same EP method to solution of higher-level, (ever
more) truncated systems of equations, starting from
Eqs.~(\ref{eq:13}) \cite{KirUSciCom,KirFractal,KirConscious}.
Dynamical fractal is different from usual, dynamically single-valued
fractals by its \emph{permanently, chaotically changing
realisations} at \emph{each} level of fractal hierarchy, which leads
to the important property of \emph{dynamic (autonomous)
adaptability} and includes \emph{any} kind of emerging structure.

Quantitative expression of dynamic adaptability takes the form of
\emph{huge efficiency growth} of unreduced many-body interaction
with respect to its unitary models. The unreduced system efficiency
$P_{{\rm{real}}}$ is determined by the link combination number in
the multivalued fractal hierarchy
\cite{KirBlindTech2,KirBlindTech4,KirNano,KirConscious,KirCommNet}:
\begin{equation}\label{eq:18}
P_{{\rm{real}}}  \propto N! \simeq \sqrt {2{\rm{\pi }}N} ({N
\mathord{\left/
 {\vphantom {N e}} \right.
 \kern-\nulldelimiterspace} e})^N  \sim N^N  \propto C\ ,
\end{equation}
where the number of links $N$ is very large itself. Unitary
(regular, sequential) dynamic efficiency grows only as $N^\beta \ll
P_{{\rm{real}}}$ ($\beta  \sim 1$). It is this huge efficiency
advantage that explains the such ``magic'' \emph{qualities} in
higher-complexity systems (very large $N$) as \emph{life,
intelligence, consciousness, and sustainability}. Obtained at the
expense of \emph{irreducible dynamic randomness}, these
\emph{causally derived} properties are \emph{indispensable} for the
\emph{correct} analysis of planetary life and civilisation dynamics.

Further development of the universal concept of complexity includes
unified classification of all observed dynamic regimes and
transitions between them
\cite{KirUSciCom,KirBlindTech2,KirSelfOrg,KirCommNet}. The limiting
regime of \emph{uniform, or global, chaos} is obtained for
\emph{comparable} interaction parameters (characteristic
frequencies). If they differ essentially, one gets the opposite case
of \emph{dynamically multivalued self-organisation, or
self-organised criticality (SOC)}, where rigid, low-frequency
components \emph{confine} a fractal hierarchy of similar, but
\emph{chaotically changing} realisations of high-frequency
components. This case \emph{unifies} the \emph{essentially
extended}, realistic and multivalued (internally chaotic) versions
of usual, dynamically single-valued ``self-organisation'' (that in
reality does \emph{not} describe any \emph{new, explicit} structure
\emph{emergence}), SOC, fractality, ``synchronisation'', ``chaos
control'', and ``mode locking''. We obtain also a rigorously derived
and universal \emph{criterion} of transition from SOC to the uniform
chaos, occurring around the main frequency resonance, which reveals
the true meaning of the ``well-known'' phenomenon of resonance
\cite{KirUSciCom,KirBlindTech2,KirSelfOrg,KirCommNet}. When the
frequency ratio, or ``chaoticity parameter'', grows from small
values for a quasi-regular SOC regime to unity in the global chaos
case, system behaviour follows a gradual (though uneven) change
towards ever less ordered patterns, reflecting the observed
diversity of dynamical structures.
\subsection{Universal symmetry of complexity and evolution law}\label{subsec:Evolution}
The major feature of \emph{explicit structure creation} includes
emerging elements of \emph{dynamically discrete, or quantized,
space} (structure) and \emph{irreversibly flowing time} (event,
evolution). Space element, $\Delta x$, is given by realisation
eigenvalue separation, $\Delta _r \eta _i^r$, for the unreduced EP
equation (\ref{eq:10}): $\Delta x = \Delta _r \eta _i^r$. Time
element, $\Delta t$, determines the duration of the \emph{event} of
space element emergence (or realisation change) and can be estimated
as $\Delta t = {{\Delta x} \mathord{\left/ {\vphantom {{\Delta x}
{v_0 }}} \right. \kern-\nulldelimiterspace} {v_0 }}$, where $v_0$ is
the signal propagation speed in the component structure. A universal
\emph{integral} measure of complexity is given by \emph{action},
$\mathcal A$, whose increment is \emph{independently} proportional
to $\Delta x$ and $\Delta t$
\cite{KirUSciCom,KirBlindTech2,KirUSymCom,KirConscious}: $\Delta
\mathcal{A} = - E\Delta t + p\Delta x$, where the coefficients $E$
and $p$ are identified as generalised system \emph{energy (mass)}
and \emph{momentum}. They represent thus universal
\emph{differential} measures of complexity:
\begin{equation}\label{eq:19}
E =  - \frac{{\Delta \mathcal{A}}}{{\Delta t}}\left| {_{x  =
{\rm{const}}} } \right.,\ \ \ p = \frac{{\Delta
\mathcal{A}}}{{\Delta x}}\left| {_{t  = {\rm{const}}} } \right..
\end{equation}
Due to its irreversible (chaotic) character, any real interaction
process can be described as \emph{transformation and conservation
(symmetry) of complexity}, where the potential (hidden) form of
complexity, or \emph{dynamic information} $I$, is transformed into
the unfolded (explicit) form of \emph{dynamic entropy} $S$, so that
their sum, the \emph{total system complexity} $C = I + S$, remains
\emph{unchanged}, $\Delta C = 0,\ \Delta I = - \Delta S$. Although
both dynamic information and entropy are expressed in units of
action, the latter corresponds rather to dynamic information
decreasing during system complexity development:
\begin{equation}\label{eq:20}
\Delta I = \Delta \mathcal{A} =  - \Delta S < 0\ .
\end{equation}
Dividing Eq.~(\ref{eq:20}) by $\Delta t\left| {_{x  = {\rm{const}}}
} \right.$, we obtain differential expression of the symmetry
(conservation) of complexity and \emph{universal dynamic/evolution
equation} in the form of generalised Hamilton-Jacobi equation:
\begin{equation}\label{eq:21}
\frac{{\Delta \mathcal{A}}}{{\Delta t}}\left| {_{x  =  {\rm{const}}}
} \right. + H\left( {x,\frac{{\Delta \mathcal{A}}}{{\Delta x}}\left|
{_{t  = {\rm{const}}} } \right.,t} \right) = 0\ ,
\end{equation}
where the \emph{Hamiltonian}, $H = H(x,p,t)$, expresses the
differential complexity-entropy, $H = \left( {{{\Delta S}
\mathord{\left/ {\vphantom {{\Delta S} {\Delta t}}} \right.
\kern-\nulldelimiterspace} {\Delta t}}} \right)\left| {_{x = {\rm
const}} } \right.$. The \emph{dynamic quantization} procedure
relates complexity-action increment to that of the \emph{generalised
wavefunction (or distribution function)} $\Psi$, describing
specific, ``disentangled'' system state \emph{during} its chaotic
jumps between realisations, and transforms Eq.~(\ref{eq:21}) to the
\emph{universal Schr\"{o}dinger equation}
\cite{KirUSciCom,KirBlindTech2,KirUSymCom,KirConscious}:
\begin{equation}\label{eq:22}
{\mathcal A}_0 \frac{{\Delta {\Psi} }}{{\Delta t}} = \hat H\left(
{x,\frac{\Delta }{{\Delta x}},t} \right){\Psi}\ ,
\end{equation}
where ${\mathcal A}_0$ is a characteristic action value by modulus
(equal to Planck's constant at the lowest, quantum levels of
complexity) and the Hamiltonian operator, $\hat H(x,p,t)$, is
obtained from the Hamiltonian $H(x,p,t)$ by causal quantization.
While the symmetry of complexity unifies and extends \emph{all}
(correct) laws and ``principles'' of the unitary science, the
Hamilton-Schr\"{o}dinger equations,
Eqs.~(\ref{eq:21})--(\ref{eq:22}), connected by causal quantization,
unify and extend all particular (model) dynamic equations
\cite{KirUSciCom,KirBlindTech2,KirUSymCom,KirConscious}.

\begin{figure}
\centerline{\includegraphics[width=11cm]{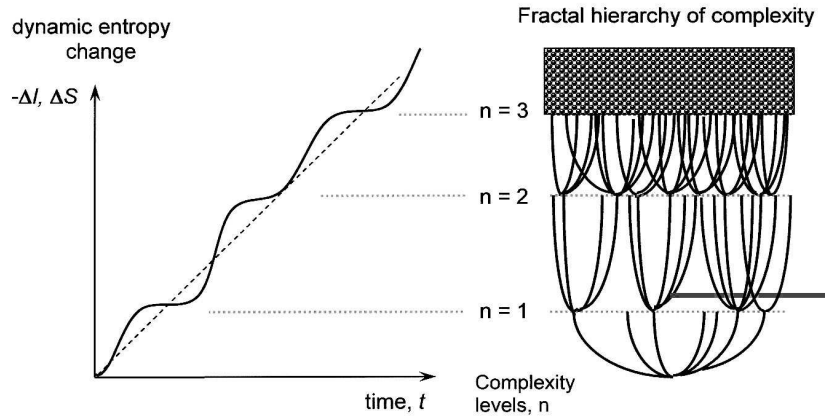}}
\caption{Scheme of universal system development by transformation of
its (decreasing) complexity-information ($I$) into (increasing)
complexity-entropy ($S$).} \label{EntropyGrowth}
\end{figure}

The key implication of the symmetry of complexity is that it
provides the \emph{universal meaning, dynamics, and measure of any
system existence, evolution, and progress}, in the form of
\emph{complexity development} (internal transformation from dynamic
information into dynamic entropy) as a result of its
\emph{conservation}, which gives a well-specified solution to such
``difficult'' and ``ambiguous'' problems as purpose, or sense of
history, meaning of life, objective understanding (and
\emph{constructive creation}) of the future, etc. Due to the
internal chaoticity of \emph{any} real (even externally ``regular'')
system, every structure emergence process corresponds to
\emph{growth of complexity-entropy, or chaoticity}, which resolves
the long-standing contradiction between entropy growth law and
visible order increase in structure creation processes. Another
complexity development feature is that due to the \emph{unreduced}
interaction dynamics (``everything interacts with everything'') it
has a \emph{dynamically discrete}, step-wise character
\cite{KirUSciCom}. The hierarchic structure creation and complexity
development process is shown schematically in
Fig.~\ref{EntropyGrowth}. A sufficiently big step of
complexity-entropy growth can be described as \emph{generalised
phase transition} to the superior \emph{level of complexity} with a
\emph{qualitatively} different kind of structure and dynamics.

\begin{figure}
\centerline{\includegraphics[width=11cm]{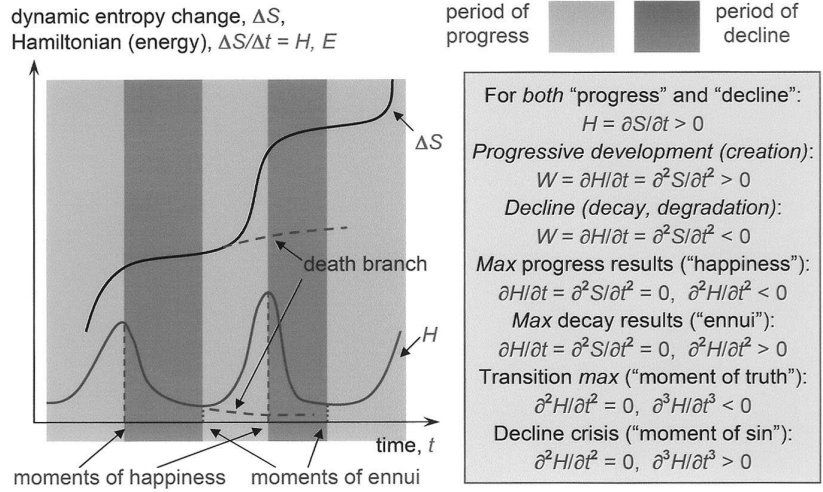}}
\caption{Periods of system progress, decline, and transitions
between them rigorously specified in terms of dynamic entropy change
$\Delta S = - \Delta {\mathcal A}$, generalised Hamiltonian $H =
{\partial S}/{\partial t}$, or energy $E = - {\partial {\mathcal
A}}/{\partial t} = H$, and higher complexity-entropy/action
derivatives.} \label{ProgressDecline}
\end{figure}

Let us consider complexity development stages in more detail,
Fig.~\ref{ProgressDecline}, in view of further application to
(modern) \emph{civilisation development}. First of all, we can
rigorously define periods of \emph{progress} (accelerated
complexity-entropy growth) and \emph{decline} (relative stagnation
of complexity development) constituting respectively the steep rise
and plateau (saturation) of each discrete step of system complexity
development. Whereas entropy $S$ can only grow for both progress and
decline, $H = {\partial S}/{\partial t} = - {\partial {\mathcal
A}}/{\partial t} = E > 0$, \emph{acceleration} of dynamic entropy
growth, or the \emph{power of development}, $W = {\partial
H}/{\partial t} = {\partial ^2 S}/{\partial t^2}$, is \emph{positive
for progress} (creative development), $W = {\partial ^2 S}/{\partial
t^2} > 0$, and \emph{negative for decline} (decay, degradation), $W
= {\partial ^2 S}/{\partial t^2} < 0$. Points of \emph{inflection}
of the entropy growth curve, ${\partial H}/{\partial t} = {\partial
^2 S}/{\partial t^2} = 0$, separate adjacent periods of progress and
decline and correspond, at the same time, to maximum (final)
progress results (``point of \emph{happiness}''), ${\partial
H}/{\partial t} = 0,\ {\partial ^2 H}/{\partial t^2} < 0$, and
maximum decay (``point of \emph{sadness/ennui}''), ${\partial
H}/{\partial t} = 0,\ {\partial ^2 H}/{\partial t^2} > 0$. One can
also define the moment of objective \emph{culmination} of a
step-wise complexity jump, progressive transition climax, or the
\emph{moment of truth} as the point of inflection of rising $H(t)$
curve, ${\partial ^2 H}/{\partial t^2} = 0,\ {\partial ^3
H}/{\partial t^3} < 0$, after which progressive complexity-entropy
upgrade becomes eminent and irreversible. In a similar way, a
critical inflection point within the period of decline, or the
\emph{moment of sin}, ${\partial ^2 H}/{\partial t^2} = 0,\
{\partial ^3 H}/{\partial t^3} < 0$, marks the definite
establishment of stagnation and decay. Whereas the points of
happiness and sadness separate the periods of progress and decline
as such, the moments of truth and sin, situated within (around the
middle of) progress and decline periods, separate the intervals of
maximum \emph{subjective} perception of their \emph{results} within
the system. We can see that such ``vague'' and ``inexact'' notions
as happiness, sorrow, and ``psychological crises'' between them are
provided with unambiguous and \emph{rigorous} definitions within the
\emph{unreduced} science of complexity (one should not forget, of
course, the whole underlying \emph{interaction analysis},
Sec.~\ref{subsec:UnredInteract}) \cite{KirUSciCom}.

Note that \emph{partial} time derivatives in the above definitions
of system evolution stages correspond to \emph{external} observation
over system development from a (generalised) \emph{reference (rest)
frame}. If now an observer is situated \emph{within} the developing
system, he will see similar development stages, but appearing on a
different, ``internal'' time scale and determined by the respective
\emph{total} time derivatives, such as (generalised)
\emph{Lagrangian} $L = - {dS}/{dt}$ \cite{KirUSciCom,KirUSymCom}.
The difference between those two time flows constitutes the
\emph{causal, complex-dynamic} basis of \emph{generalised special
relativity} effects emerging at \emph{all} levels of dynamics, from
quantum particle motion to civilisation development
\cite{KirUSciCom}.

Note finally that progressive transition to superior level of
complexity can be replaced by another development branch, the
``death branch'' of purely destructive degradation of existing
system structures, without qualitatively new, ``progressive''
structure emergence (Fig.~\ref{ProgressDecline}). This scenario
becomes real when the stock of complexity-information of the driving
interaction process is exhausted or when further complexity
development is seriously blocked in a deep impasse (``wrong way'').
In the first case one deals with the generalised complex-dynamical
system \emph{death}, which is now \emph{rigorously} defined
\cite{KirUSciCom} and inevitable (for a closed system) because of
the finite quantity of dynamic information, whereas in the second
case one has a \emph{bifurcation of development}, where \emph{both}
progressive transition to a higher complexity level \emph{and}
destructive degradation can happen with certain, dynamically
determined \emph{probabilities} (see Eq.~(\ref{eq:17})).
\section{Sustainability transition as the revolution of complexity}\label{sec:Transition}
\markright {Sect. \ref{sec:Transition}: Sustainability transition as
the revolution of complexity}
\subsection{Modern bifurcation of civilisation development: Causal Apocalypse now}\label{subsec:TodayBifurc}
We can apply now the unified development theory from the previous
section to modern civilisation development, including its recent
past and forthcoming future. Observed features analysis shows that
modern civilisation, suitably represented by its advanced,
``locomotive'' parts, is situated in the vicinity of the last
``point of sadness (ennui)'' (Fig.~\ref{ProgressDecline}) and maybe
already slightly outside of it in the direction of a probable
complexity-growth step (but well before its ``moment of truth'').
That modern world position at the beginning of emerging inflection
of $H(t)$ curve after its deep minimum (development saturation) is
supported by a variety of clearly observed ``ends'', such as End of
History, Science, Art, Religion, etc. (e.g.
\cite{Rees,Soros,EndScience,EndHistory}), appearing as a stable
absence of true novelty emergence (events) and pronounced
degradation of existing structures \cite{KirUSciCom}. In view of the
close ``death branch'' beginning (Fig.~\ref{ProgressDecline}), we
get to the \emph{great bifurcation of development} into the death
branch of pure destruction and transition to a superior level of
civilisation complexity. Taking into account the huge,
\emph{ultimately complete} scale of all ``ends'' involved, we can
say that we deal here with the rigorously substantiated version of
Apocalyptic ``End of the World'', Doomsday, etc. appearing in
reality as that major development bifurcation into two main branches
of ``(system) death'' and ``(new) life'', where the latter emerges
by transition to a \emph{qualitatively} higher complexity level of
the \emph{whole} civilisation dynamics
\cite{KirUSciCom,KirSustTrans}. The latter change can also be
designated as \emph{Revolution of Complexity}, or
\emph{sustainability transition} (see
Secs.~\ref{subsec:NewScience}--\ref{sec:Conclusion}).

Particular, \emph{practically important} results of this
\emph{rigorously derived} development concept are specified below
(Secs.~\ref{subsec:NewScience}--\ref{subsec:NewSettle}). The
\emph{causally complete} nature of the underlying interaction
analysis (Sec.~\ref{sec:USciCom}) leaves practically no hope that
the observed bifurcational, ``Apocalyptic'' state of modern
civilisation can be avoided by usual, ``smooth'' amelioration of
life conditions, often subjectively privileged by prosperous,
``leading'' civilisation components (e.g. within standard,
``protective'' ecological actions,
Secs.~\ref{subsec:NewProd}--\ref{subsec:NewSettle}). Dynamic entropy
growth \emph{cannot} stop, but the failure to follow the strongly
growing, \emph{qualitative} development branch at the \emph{current
specific moment} will \emph{inevitably} leave civilisation on the
death branch of irreversible destruction. We see that the
\emph{causally complete understanding} of \emph{unreduced, unified}
civilisation dynamics within the universal science of complexity
provides the unique and \emph{vitally} important basis for the
\emph{scientifically exact, rigorous futurology}
(Sec.~\ref{sec:Conclusion}).
\subsection{The last scientific revolution}\label{subsec:NewScience}
It is convenient to start our more detailed analysis of
sustainability transition and the resulting superior complexity
level with the description of respective changes in the \emph{system
of knowledge}, the more so that the new level of complexity is
characterised by a much greater, \emph{decisive} role of a \emph{new
kind} of ordered, ``scientific'' knowledge in the whole civilisation
development.

Unitary, dynamically single-valued science approach dominating today
(and including \emph{zero-complexity imitations} of ``complexity''
and ``chaoticity'') is unable to provide consistent understanding of
\emph{any} real, \emph{dynamically multivalued} system behaviour
(Sec.~\ref{subsec:UnredInteract}), which becomes especially evident
for higher-complexity cases (strong interaction, living organisms,
intelligent behaviour, social, ecological systems, etc.). At the
same time, the \emph{purely empirical}, technological civilisation
power has attained today, \emph{for the first time in history}, the
critical threshold of the \emph{full depth} of \emph{any} real
system complexity, from quantum world (elementary particles and
fields) to the structure of life (genome and related cell processes,
ecosystems, brain processes). This \emph{effectively blind} but
quantitatively powerful, ``stupid'' technology uses the conventional
trial-and-error empiricism to \emph{strongly modify} systems whose
real dynamic complexity exceeds by far the possibilities of
zero-dimensional ``models'' of unitary science (they are still
shamelessly promoted for ``simulation'' of ultimately complex
behaviour of economic, social, and ecological systems!). The
resulting contradiction creates \emph{real and unprecedented
dangers} at \emph{all} complexity levels, from particle physics to
genetics and ecology, which are \emph{not} due to the ``risk of
science/technology'' in general (cf. \cite{Rees}), but due to the
\emph{specific, artificial limitation} of the \emph{unitary} science
paradigm and results \cite{KirUSciCom}.

Transition to another, \emph{causally complete} kind of knowledge is
therefore \emph{urgently needed} today and the failure to perform it
will \emph{inevitably} lead to destructive consequences, as the
probability of successful empirical ``guess'' or unitary
``simulation'' of the huge power of real system complexity (see
Eq.~(\ref{eq:18})) is very close to zero. It is clear that the new,
\emph{practically efficient} knowledge can only be based on the
detailed understanding of the \emph{unreduced} interaction process
underlying any real system dynamics, which leads directly to the
dynamic multivaluedness paradigm \cite{KirQuChaos,KirChannel} and
universal science of complexity (Sec.~\ref{sec:USciCom})
\cite{KirUSciCom}. Being thus indispensable for real problem
solution already at the existing level of development, the unreduced
science of complexity becomes \emph{unified and unique basis} for
realisation of sustainability transition and resulting superior
level of civilisation complexity. It is this, ultimately complete
and realistic kind of knowledge that can form a practical basis for
the ``\emph{society based on knowledge}'' at the superior complexity
level. It is clear also that imitations of the unitary ``science of
complexity'' can only be harmful because of their \emph{biggest
possible}, dynamically single-valued simplification of real system
dynamics. Practical \emph{organisation of science} should follow the
corresponding \emph{qualitative} change towards a much more liberal,
decentralised and adaptable system with \emph{emergent} structure
\cite{KirUSciCom,KirBlindTech3,KirBlindTech2}.

The essential extension of science content, role, and organisation
constitutes thus a major part of the forthcoming  Revolution of
Complexity. The latter can be considered, in this sense, as the
\emph{last} ``scientific revolution'' of the kind described by
Thomas Kuhn \cite{Kuhn}, since the unreduced science of complexity
realises the \emph{intrinsically complete, permanently creative}
kind of knowledge, devoid of \emph{antagonistic} fight between
``paradigms'' and people (which originates, as it becomes clear now,
from the \emph{specific}, strongly \emph{imitative} nature of the
\emph{unitary}, ``positivistic'' science, rather than scientific
knowledge in general).
\subsection{Complexity-increasing production: Growth without destruction and the universal criterion of progress}\label{subsec:NewProd}
Modern industrial production leads to evident and rapid degradation
of environment and life quality, and therefore \emph{cannot} provide
\emph{long-term progress}. As such progress is a \emph{necessary}
condition for planetary civilisation \emph{existence}, one is
brought to the idea of \emph{sustainable development}. However, the
self-protective approach of the current system tends to the tacit
assumption that sustainability can be attained by gradual
``purification'' of production methods, without major, qualitative
change of the dominating industrial mode as such
\cite{ChangEarth,Vellinga,Lillo,Jager}.

The unreduced interaction analysis of the universal science of
complexity rigorously shows, first of all, that the latter hope is
totally vain and sustainability \emph{cannot} be attained within the
current way of production, irrespective of the details, simply
because it is invariably reduced to \emph{destruction} of
complexity, i.e. transformation of higher-complexity structures into
lower-complexity ones \cite{KirUSciCom,KirSustTrans}. We also use
here the rigorous and universal \emph{definition and criterion of
progress} as \emph{optimal growth of complexity-entropy} according
to the system development curve
(Figs.~\ref{EntropyGrowth},~\ref{ProgressDecline}). At the modern
moment of maximum/ending stagnation (Sec.~\ref{subsec:TodayBifurc})
civilisation progress can only proceed by self-amplifying
complexity-entropy growth towards its superior level, without which
the system will inevitably follow the death branch of catastrophic
destruction.

This criterion of progress can be provided with exact formulation by
recalling that transition to superior complexity level acquires a
well-defined character after the ``moment of truth'', or
Hamiltonian/Lagrangian inflexion point, where the second time
derivative of Hamiltonian/Lagrangian changes sign from plus to minus
(Fig.~\ref{ProgressDecline}). The ensuing criterion of progress (in
its ``internal'' version expressed by total time derivatives) is
\begin{equation}\label{eq:23}
\frac{{d^3S}}{{dt^3}} < 0\ ,\ \ \textrm{or}\ \ \
\frac{{d^2L}}{{dt^2}} > 0\ ,
\end{equation}
where $L = - {dS}/{dt}$ is the system Lagrangian
(Sec.~\ref{subsec:Evolution}) \cite{KirUSciCom,KirUSymCom}. Note
that progressive \emph{development} thus defined overlaps with both
\emph{periods} of progress and decline defined before
(Sec.~\ref{subsec:Evolution}) and includes their ``best'' parts of
\emph{essential}, self-amplifying growth of complexity-entropy (even
though its \emph{rate}, ${dS}/{dt}$, decreases within the beginning
period of decline). A \emph{narrow} understanding of ``definite''
progress would include only progressive development part
\emph{within} the period progress, ${d^2L}/{dt^2} > 0,\ {dL}/{dt} <
0$, while the \emph{whole} progressive development can also be
designated by the condition ${dS}/{dt} \gg
({dS}/{dt})_{{\rm{death}}}$, where $({dS}/{dt})_{{\rm{death}}}$ is
the maximum entropy growth rate for the death branch (or its minimum
value for the decline period).

Impossibility of sustainable development at the current complexity
level follows from the generalised entropy growth law: any, even
``ecologically correct'' production of the current, industrial way
can at best only minimise the inherent complexity destruction
(entropy growth), but can never reduce this high enough minimum to
values around zero. But the \emph{same} entropy growth law underlies
\emph{genuine sustainability} at the \emph{superior} complexity
level, after the key transition to \emph{complexity-increasing
production} methods and technologies. That's why it is called
\emph{sustainability transition} (Sec.~\ref{subsec:TodayBifurc}).
Indeed, in this case the inevitable complexity-entropy growth takes
the form of intrinsically \emph{progressive} creation of \emph{ever
more complex} structures (``period of progress'' in
Fig.~\ref{ProgressDecline}), as opposed to a ``period of decline''
where entropy growth is dominated by destruction of previously
created structures. It means that the criterion of progressive
development, Eq.~(\ref{eq:23}), remains practically \emph{always
valid} after the sustainability transition, and very short periods
of formal ``decline'' are determined by decreasing, but \emph{high}
rate of entropy growth, ${d^2S}/{dt^2} < 0,\ {dS}/{dt} \gg
({dS}/{dt})_{{\rm{death}}}$, \emph{within} progressive development,
${d^3S}/{dt^3} < 0$.

Realistic basis for production sustainability is due to
\emph{complexity creation} and \emph{complexity-based} kind of
technology, where the unreduced complexity-entropy of \emph{all}
production results should be \emph{essentially greater}, than that
of the initial system configuration. An important example is
provided by \emph{irreducibly complex dynamics} of \emph{realistic}
sources of pure energy from \emph{nuclear fusion reactions} (in its
both ``hot'', less sustainable and ``cold'', more prospective
versions), demonstrating the \emph{unified, multi-level} structure
of the Complexity Revolution.

Contrary to popular ideas about industrial production, its
complexity-killing features are \emph{not} due to \emph{massive} use
of man-made \emph{machines} as such, but due to a certain,
``unitary'' way of using certain kind of machinery. Those
particular, complexity-reducing tools and methods are closely
related to \emph{specific organisation} of usual industrial
production characterised by explicitly \emph{reduced dynamic
complexity} (tendencies of unification, regularity, etc.).
Correspondingly, the new, intrinsically sustainable production at
the superior complexity level should be organised in a qualitatively
different way dominated by the permanently developing, hierarchic,
distributed ``ecosystem'' of \emph{dynamically} connected, generally
small units of \emph{individually} structured production (they
certainly can form \emph{loose, dynamic} associations at higher
ecosystem levels that will replace modern inefficient and decadent
corporate monsters). It becomes evident that such
complexity-increasing production organisation and content is
inseparable from the accompanying \emph{personal progress} of human
complexity, i.e. the level of \emph{consciousness}
\cite{KirConscious} (see also
Secs.~\ref{subsec:NewSocial}--\ref{sec:Conclusion}).
\subsection{From unitary to harmonical social structure:\\
Emerging order without government}\label{subsec:NewSocial}
Due to \emph{holistic} dynamics of unreduced interaction
\cite{KirUSciCom}, sustainability transition involves a qualitative
change of \emph{social structure and dynamics}. In order to specify
this change, we show first that social structure of the current
complexity level, including \emph{all} known (modern and ancient)
social and political regimes, constitutes a single kind of order
called \emph{Unitary System} \cite{KirBlindTech3,KirSustTrans}. The
term ``unitary'' (behaviour) has a \emph{mathematically exact}
interpretation in the universal science of complexity
(Sec.~\ref{subsec:UnredInteract}) of ``dynamically single-valued''
and therefore \emph{qualitatively uniform}, regular,
zero-complexity, ``effectively one-dimensional'', sequential
(dynamics, evolution, etc.). Although any social system cannot be
strictly unitary in this rigorous sense, the Unitary System of
social structure is close to it because it is a rigid, centralised
system of \emph{preferably} regular (controlled) dynamics that can
change essentially (usually just to its another \emph{version}) only
by way of \emph{destructive} ``revolution''. Such unitary social
order includes \emph{all} previously known social systems (usually
considered to be very different), such as any totalitarian,
democratic, or meritocratic political structure. Correspondingly,
social structure resulting from sustainability transition should
differ \emph{qualitatively} from \emph{any} of these, including
allegedly the ``best possible system'' of modern democracy (as well
as any meritocracy).

We call the social organisation type of that qualitatively superior,
higher-complexity level \emph{Harmonical System}
\cite{KirBlindTech3,KirSustTrans}. Contrary to any version of
Unitary System, the Harmonical System has the \emph{emergent},
intrinsically \emph{creative}, permanently developing kind of social
order whose origin resembles that of the free market
\emph{economical} structure, but encompasses now the whole
civilisation structure. It is dominated by a system of interacting,
independent units similar to those of the complexity-increasing
production structure (Sec.~\ref{subsec:NewProd}), but including
\emph{all} spheres of activity. \emph{Global} system dynamics is
monitored by the \emph{same} kind of \emph{independent}, interactive
units, very different from any unitary ``government'' (or even
``non-government organisations'', NGO) in that they are
\emph{forced} to produce \emph{explicitly useful} services,
\emph{compete} with each other and bear \emph{individual,
well-specified responsibility} for their results (similar to
\emph{small} enterprises within market economy). Any loose
associations of such units, as well as ``high councils'', may exist,
but only as far as they are needed and without any formal power
\emph{exceeding} that of \emph{emergent} actions of independent
enterprises (including various ``forces of order'').

Note that some seeds of such emergent social order may exist within
modern ``developed'' version of unitary democracy, but any its most
``liberal'' version or component (like NGO) is severely limited by
the \emph{imposed} rigid, formal (``obligatory''), centralised power
ensuring the status of unrealistic dream for any true liberty (=
unreduced, natural, progressive development). The Harmonical System
of \emph{emergent order} realises what is considered as impossible
by the conventional, unitary democracy, a \emph{qualitatively
higher} kind of liberty and ``democratic'' order obtained
\emph{without} any ``majority vote'' (always manipulated by
``minority games''). This ``miracle'' becomes possible only at the
described superior level of civilisation complexity realised by
unreduced interaction of independent units pervading \emph{all}
spheres of activity
(Secs.~\ref{subsec:NewScience}--\ref{subsec:NewSettle}).

The harmonical social order has \emph{intrinsically progressive}, or
\emph{sustainable} structure due to \emph{permanent},
non-antagonistic and \emph{essential} complexity development in the
sense of our rigorously defined progress
(Sec.~\ref{subsec:NewProd}). The very character of civilisation
development changes \emph{forever} after sustainability transition,
from painful alternation of ``stagnation'' and ``revolution''
periods to the \emph{permanent unreduced creativity} (that could
also be described as ``distributed complexity revolution''). By
contrast, the modern ``developed'' unitary democracy, apparently
repeating respective periods of ancient civilisation development,
represents not the ``best possible'' social system (according to its
own praise, thoroughly maintained by self-privileged ``powers that
be''), but rather the \emph{definite end}, generalised
complex-dynamical \emph{death-equilibrium} \cite{KirUSciCom} of the
Unitary System as such, in \emph{any} its version, followed
inevitably by either sustainability transition to a superior
complexity level of Harmonical System, or irreversibly destructive
death branch (Sec.~\ref{subsec:TodayBifurc},
Fig.~\ref{ProgressDecline}). In fact, this ``final'', decadent,
\emph{equilibrium} character of unitary democracy, clearly seen
today, does result from its \emph{highest possible} development of
the \emph{unitary} kind of social structure that does not need to
be, however, its \emph{only} possible kind and actually represents
the \emph{simplest}, basically ``tribal'' (imposed,
compulsion-based) kind of social order. The latter becomes
insufficient today just \emph{due to} the ultimately high
development of the industrial Unitary System creating
self-amplifying, and therefore \emph{insurmountable}, dynamic
barriers to its own progress.
\begin{figure}
\centerline{\includegraphics[width=11cm]{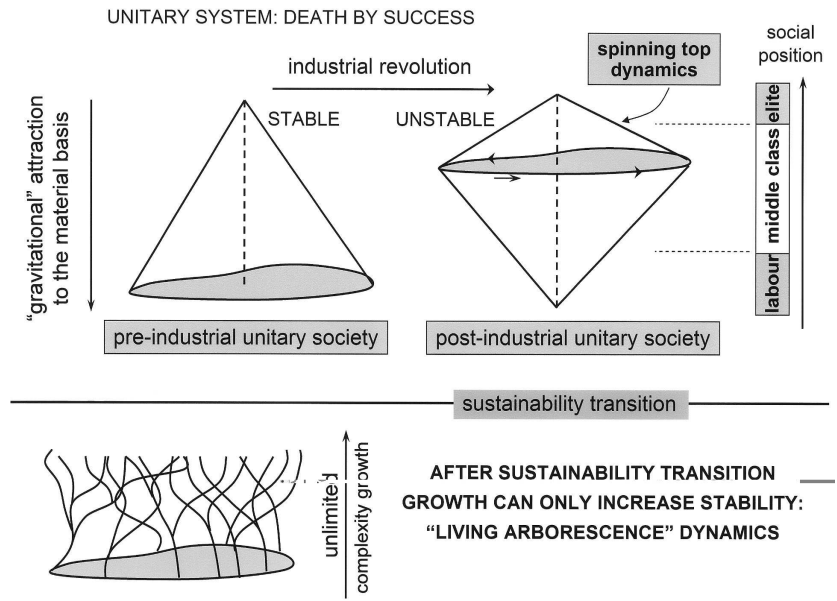}}
\caption{Schematic representation of the critical instability of the
developed (modern) Unitary System followed by globally stable
structure of the Harmonical System.} \label{CriticalInstability}
\end{figure}

The origin of modern, \emph{inevitably} emerging \emph{critical
instability} of the \emph{developed}, industrial Unitary System can
be conveniently demonstrated with the help of schematic presentation
of its social structure dynamics, Fig.~\ref{CriticalInstability}.
Pre-industrial, ``traditional'' Unitary System can be presented by a
pyramidal structure stably resting on its large base of labour
classes due to the ``gravitational attraction'' towards material
production/consumption. In the post-industrial society, the same
unitary pyramid acquires a strongly deformed, ``inverse''
(upside-down) configuration due to huge productivity growth as a
result of technological revolutions. But since the material
``gravity'' force preserves the same downward orientation, that
monstrous construction with now \emph{quantitatively} dominating
non-productive ``elitary layers'' becomes critically unstable and
can preserve its normal, ``vertical'' position only due to the
high-speed spinning motion of production-consumption cycles (similar
to spinning top stability). However, this \emph{artificially}
maintained, \emph{relative} stability has its limits, especially due
to basically \emph{dissipative}, chaotic dynamics of any social
system, which means that the unitary ``whipping top'' \emph{will}
fall in a destructive manner within a reasonably small time period
(like few tens of years). By contrast, the harmonical social
structure, shown at the bottom of Fig.~\ref{CriticalInstability} as
a distributed arborescence, does not possess any global, destructive
instability: instead, its \emph{local, creative} instability
provides sustainable progress.

Harmonical System represents thus the \emph{unique} way of
\emph{any} further progress, and in order to realise it one should
have a \emph{realistic} sustainability transition. Such realistic
transformation takes the form of generalised ``phase transition'' of
\emph{higher order}, where the \emph{qualitatively big} structure
change occurs not in the whole system volume simultaneously (as in
``first-order transitions''), but starts with \emph{small},
\emph{growing} ``seeds'' of the ``new phase'', which strongly
facilitates the transition process. The dynamics of both
sustainability transition and resulting Harmonical System can be
properly understood and monitored only with the help of the
\emph{causally complete} understanding of the unreduced science of
complexity (Sec.~\ref{subsec:NewScience}), which emphasises once
more the high role of this \emph{new kind of knowledge} in the
forthcoming development stages.
\subsection{New settlement and infrastructure}\label{subsec:NewSettle}
It is not surprising that civilisation infrastructure at the unitary
level of development, including the dynamical structure of
settlements, production and communications, reflects major features
of the Unitary System, such as high centralisation, rigidity,
development rather by destruction, pronounced tendency towards
mechanistic simplification, and the resulting urban decadence in the
phase of ``developed'' unitarity. Indeed, there is the evident
degradation to over-simplified, ``squared'' and ``smooth''
configurations and operation modes in modern infrastructures,
despite \emph{much greater} practical possibilities for their
diversity in the developed industrial technology. Whereas this
effective complexity destruction is a part of the emerging ``death''
tendency of the ending level of development
(Sec.~\ref{subsec:TodayBifurc}), it is equally evident that the
forthcoming harmonical level of complexity should be based on a
\emph{qualitatively different} type of settlement with a
distributed, decentralised, and \emph{progressively developing}
structure (see Sec.~\ref{subsec:NewProd} for the universal progress
definition).

This another kind of settlement can only be realised as a man-made
structure intrinsically and strongly \emph{submerged} into the
``natural environment'' and \emph{constructively} interacting with
it, so as to \emph{increase} complexity-entropy of the whole system.
Such sustainable civilisation structure can be described as
omnipresent, man-controlled, progressively evolving \emph{forest},
or ``natural park'', with submerged, \emph{distributed} settlement,
production and transport infrastructure, which excludes anything
closely resembling modern cities, towns, and villages, with their
\emph{centralised} structure tendency. Transport networks in such
``living'' infrastructure will be well hidden among other, more
``natural'' and complexity-bearing elements, contrary to their
domination in the unitary infrastructure. The omnipresent and
intense \emph{creation} of ``natural'', i.e. \emph{complex-dynamic}
environment, rather than its unitary ``protection'' (inevitably
failing), constitutes the essence of complexity-increasing
settlement and infrastructure dynamics. The latter correlates
directly with the complexity-increasing production mode
(Sec.~\ref{subsec:NewProd}) because it can be considered as a
specific sphere of production with strong involvement of ``human
dimensions''.

Progressively growing dynamic complexity of this new kind of
``natural'' but totally man-controlled environment and
infrastructure has a positive reverse influence upon dynamic
complexity of man's consciousness and life style. This positive
feedback loop in the man-environment system leads to a dynamic
complexity boost that can be described as \emph{Supernature} at the
level of ``environment'' structure and as (\emph{realistically}
specified) \emph{Noosphere} at the level of human consciousness
(including its individual and ``social'' aspects). Supernature can
have the same or even \emph{much greater} dynamic complexity than
the ``wild'' nature (contrary to any ``protected'' environment of
the unitary ecology), while Noosphere emerges as inseparable,
fractally structured, and progressively evolving \emph{dynamic
entanglement} (Sec.~\ref{subsec:UnredInteract}) of (superior)
consciousness and Supernature. In this sense one can say that
nature, in its new form of Supernature, should become again man's
home, at this superior, harmonical (complexity-increasing) level of
their interaction.
\section{Cosmic intelligence, future, and complexity: Concluding remarks}\label{sec:Conclusion}
\markright {Sect. \ref{sec:Conclusion}: Cosmic intelligence, future,
and complexity: Concluding remarks}
Summarising the universal science of complexity
\cite{KirUSciCom,KirBlindTech3,KirBlindTech1,KirBlindTech2,KirBlindTech4,KirSelfOrg,KirUSymCom,KirSustTrans,KirQFM,KirCosmo,KirQuChaos,KirNano,KirFractal,KirConscious,KirCommNet,KirChannel}
(Sec.~\ref{sec:USciCom}) and its application to the problems of
modern civilisation development (Sec.~\ref{sec:Transition}), one
should emphasize \emph{intrinsic unification} of \emph{causally
specified} meaning and purpose of life, future, progress, nature,
cosmos, and our destiny within the \emph{universal symmetry of
complexity} (Sec.~\ref{subsec:Evolution}), thus constituting the
\emph{practical guiding principle for civilisation development}.

Application of the unreduced science of complexity to the problem of
cosmic life and extraterrestrial civilisations shows that life
realisations in cosmos should be \emph{multiple and diverse}, while
unique civilisation existence is highly improbable: it follows
already from the basic property of \emph{dynamic multivaluedness}
(Sec.~\ref{subsec:UnredInteract}). The \emph{complexity
correspondence principle} following directly from the universal
symmetry of complexity \cite{KirUSciCom,KirBlindTech2} provides a
\emph{rigorous} basis for the statement that real, constructive
contact between different civilisations is possible if they have
similar levels of unreduced complexity (consciousness) that should
certainly be high enough for the contact at a cosmic scale.
Therefore the complexity/consciousness upgrade of a particular
civilisation of the planet Earth, which is necessary for its own
development (Sec.~\ref{sec:Transition}), can be a much more
efficient way of establishing contact with extraterrestrial
intelligence than usually applied technical means (``find an alien
within yourself'').

There is no other way to a sustainable, non-destructive future than
essential growth of civilisation complexity taking the form of
\emph{Revolution of Complexity} in all fields of human activity
(Sec.~\ref{subsec:TodayBifurc}). But since the latter is determined
by the level of consciousness that can be causally understood itself
as a high enough level of complex interaction dynamics
\cite{KirUSciCom,KirConscious}, it becomes evident that modern
\emph{bifurcation of development} is centred around that
\emph{critical consciousness upgrade}, which constitutes today the
main factor of \emph{civilisation survival}: real Future comes as a
superior level of individual consciousness. It shows the emerging
predominant role of \emph{individually} specified results of
\emph{global interaction} processes, as opposed to conventional
``mass consciousness'' effects of the unitary society at previous
development stages. In fact, only consciousness complexity
development provides the basically \emph{unlimited progress
perspective} after the objective end of the unitary history of
``hot'' events (cf. \cite{EndHistory}).

As \emph{every} future becomes \emph{uncertain} at a qualitative
transition point of modern Apocalyptic scale
(Sec.~\ref{subsec:TodayBifurc}), one should understand now
\emph{all} possible futures within a \emph{unified} vision, by
contrast to innumerable ``scenarios'' and one-dimensional unitary
interpolation ``threads'' for \emph{separate} aspects of development
that become totally inefficient and misleading just at such critical
point of ``generalised phase transition'' \cite{KirUSciCom} (cf.
\cite{VernonVinge,ChangEarth,Rees,Soros,FutureShock,WebFuture}).
Providing a unique possibility of such unified, causally complete
vision of multiple interaction processes determining civilisation
development, the \emph{universal science of complexity} constitutes
the \emph{truly scientific} basis for \emph{consistent, provably
reliable futurology} and its critically important applications to
modern development problems \cite{KirUSciCom,KirSustTrans}.

%% These index commands will not work with SWP. You must use
%% LaTeX outside SWP to format the index.

\kluwerprintindex

%% Use these commands if you are making a separate topic and author index.
%% If you use these two commands, you will not need the \kluwerprintindex
%% above.

%\printtopicindex
%\printauthorindex

%% Author index info:
%% Enter author entries like this: \anxx{Dillon\, Matt}
%% When you run LaTeX on your file you will produce filename.aut.
%%   You will need an .att file to produce the author index.
%% Sort the filename.aut file to produce filename.att:
%%    sort filename.aut > filename.att

\end{document}